\documentclass[rsi,reprint,superscriptaddress]{revtex4-1}
\usepackage{amsmath} 
\usepackage{amsthm} 
\usepackage{amssymb}	
\usepackage{graphicx} 
\usepackage{enumerate}
\usepackage{wrapfig}
\usepackage[dvips,letterpaper,margin=1in,bottom=1in]{geometry}
\usepackage{color}
\usepackage{tensind}
\tensordelimiter{`}
\whenindex{a}{\alpha}{} \whenindex{b}{\beta}{} \whenindex{g}{\gamma}{} \whenindex{d}{\delta}{} \whenindex{e}{\varepsilon}{} \whenindex{m}{\mu}{} \whenindex{n}{\nu}{} \whenindex{l}{\lambda}{} \whenindex{k}{\kappa}{} \whenindex{r}{\rho}{} \whenindex{s}{\sigma}{} \whenindex{t}{\tau}{}
\usepackage{todonotes}
\usepackage{subcaption}
\usepackage{setspace}
\usepackage{comment}
\usepackage{cancel}

\usepackage{hyperref}
\usepackage[all]{hypcap}
\makeatletter 

\let\baraccent=\= 
\renewcommand{\=}[1]{\stackrel{#1}{=}} 
\renewcommand{\deg}{^{\circ}} 

\newcommand{\<}{\left <}
\renewcommand{\>}{\right >}

\theoremstyle{definition}

\theoremstyle{remark}

\newcommand{\um}{$ \mathrm{\mu} $m}

\newcommand{\FeGa}{Fe$ _{1-x} $Ga$ _{x} $}

\begin{document}
	\title{Real-time MOKE microscopy made simple}

	\author{Pavel Chvykov}
	\author{Vladimir Stoica}
	\author{Roy Clarke}
	\affiliation{University of Michigan}

	\begin{abstract}
		
		We present a simple and effective instrument for simultaneous real-time imaging and hysteresis of the anisotropic magnetic domain dynamics in thin films using the Magneto-Optical Kerr Effect (MOKE). We furthermore illustrate that magnetic imaging allows a more accurate interpretation of the magnetization reversal processes than the conventional hysteresis characterization. In particular, we present a case where the onset of a double-step reversal observed in imaging remains invisible in the spatially integrated hysteresis loops. When complemented by precise tuning of the external magnetic field orientation, our system reveals the singular anisotropic variations of the domain dynamics near the hard-axis in epitaxial thin films, thus shedding light on the reported, but as yet unexplained, hard axis coercivity behavior.
		
	\end{abstract}
	\maketitle
	\section{Introduction}
	
	The study of magnetism in thin-films and related nanostructures has seen an increasing popularity over the past decade, as these topics form the core of fundamental and applied research in spintronics \cite{spintronics}. Nonetheless, achieving high spatio-temporal resolution in the imaging of the magnetic domain structures is an intricate matter because magnetic domain imaging instruments are usually complex and expensive. For example, prominent magnetic imaging techniques offering nanoscale spatial resolution include the Magnetic Force Microscopy (MFM)\cite{MFM, MFM_crossover}, Scanning Electron Microscopy with Polarization Analysis (SEMPA)\cite{SEMPA, SEMPA2010}, and Magnetic Transmission X-ray Microscopy (MTXM)\cite{MTXM, MTXM2010}. Despite their distinct advantage of high spatial resolution, these techniques require intricate setups and offer limited flexibility in, e.g., quick adjustment of field of view and resolution, while the former two techniques cannot be easily be implemented for time-resolved studies \cite{domains_book}. 
	
	Magneto-Optic Kerr Effect (MOKE) microscopy using visible light sources is another common imaging technique known for its simplicity and flexibility, while also being capable of femtosecond time and diffraction-limited spatial resolutions. \cite{old_MOKE, MOKE_rever_anis, fs_MOKE} However, this method generally suffers from a weak signal (defined as the imaging contrast between spin orientations), especially if time-resolved imaging is needed, which limits the use time-averaging or scanning-type setups for signal-to-noise ratio improvement purposes. The most common MOKE microscopy setups either employ MOKE in the polar geometry, giving larger sensitivity compared to the longitudinal or transverse geometries, or use the help of Magneto-Optic Imaging Films (MOIF) for signal enhancement \cite{MOIF1}. However, the polar MOKE geometry can only record the magnetization component orthogonal to the film surface\cite{PolarMOKE}, while imaging the more common in-plane magnetization with MOIF \cite{MOIF2, MOIF_new} requires a magneto-optical transducer placed directly on top of the film surface, thus causing undesirable interference with the sample. Further, as the MOIF materials are not commercially available, they can be difficult to obtain.
	
	In this paper, we show that an effective MOKE microscope can be easily assembled using basic components available in most optics laboratories. Moreover, with our experimental setup we could perform simultaneous studies of vector-resolved magnetic hysteresis and real-time-resolved longitudinal MOKE imaging of magnetic domain dynamics in thin films. The imaging of the transverse magnetization component can also be achieved through a straightforward system reconfiguration, as can the more common polar geometry. We demonstrate that when complemented by high angular resolution in manipulating the sample orientation with respect to the external magnetic field, we can observe the fine details of the anisotropic behavior of magnetic domains during the magnetization reversal. In particular, we study the reversal dynamics on the hard-axis coercivity spike -- a sadden increase in the coercive field when the magnetic field direction is oriented along the hard axis, which was observed in \cite{spike2003,spike2011,MOKE90deg&spike} from hysteresis, but the domain structure could not be previously imaged with time-resolution. Finally, we show that key physical information about the reversal can be missed by the conventional hysteresis measurements, but is picked up in imaging, and thus simultaneous imaging and hysteresis data acquisition are essential for interpreting the magnetization dynamics.

	\section{The Experimental Setup}
	
\begin{figure}
		\centering
		\includegraphics[width=0.35\textwidth]{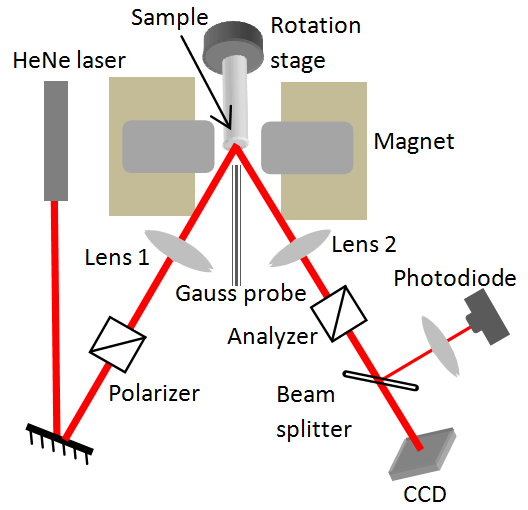}
		\caption{Experimental arrangement for observation of anisotropic reversal dynamics, including the simultaneous detection of spatially integrated hysteresis with a photodiode and with MOKE microscopy with a CCD}
		\label{fig:setup}
	\end{figure}

	Our setup is schematically presented in fig. \ref{fig:setup}. A 15 mW HeNe laser beam passes through a polarizer and is directed onto the sample by lens 1 for illumination. We observe that the size and quality of the illumination spot here are not crucial, as long the field of view is illuminated uniformly and with sufficient brightness. A lens is used for imaging the sample surface onto a CCD camera with 12 bit dynamic range, acquiring 30 frames per second with 0.1 to 1 ms variable exposure time. After the collection lens, the beam passes through an analyzer. Although the laser beam divergence does affect the analyzer performance, as the incident angle deviates from normal, this decline is insignificant and does not pose problems for the imaging contrast. A beam splitter is then used for directing the beam to both the CCD and a commercial photodiode (Thorlabs PDA10A), for which the data acquisition is performed in real time using a USB digitizer (National Instruments USB-6221). The photodiode simply provides the averaged intensity from the imaged spot, but with a much higher temporal resolution for faster and more precise hysteresis measurements than those possible with the CCD. The sample is mounted on a motorized rotation stage with $ 0.01\deg $ angular resolution and with the plane of rotation being parallel to the film surface and to the external field $ H $ applied by an electromagnet (GMW 3470). The magnet is driven by a sinusoidal current from its power supply (Kepco BOP 50-8M) seeded by a function generator or, with more flexibility, by a PC. The system provides a maximum field of ~10 kOe with a field uniformity across the sample of ~2\%. The magnetic field is measured by a Gauss probe next to the sample and all data processing occurs in real time following the acquisition on a PC using a USB data acquisition board. A LabView software routine was programmed for experiment control, which uses the Gauss-probe signal to trigger the hysteresis loop measurements, and appropriately averages typically between 1 and 30 hysteresis loops, while further plotting the loops in real-time to correlate the photodiode MOKE signal relative to the Gauss-probe-measured field. Additionally, the software routine numerically analyses the measured loops to find the parameters of interest, such as coercivity, remanence and the magnetic field at saturation, and plots these parameters live during the measurement. 
	
	In our system, for simplicity and flexibility, the sample was mounted using adhesive tape to the end of the rotation stage mount (the drift of the sample orientation settled within a few minutes after taping and did not cause problems). Due to the small spacing between the poles of the electromagnet, which was necessary to apply larger magnetic fields, the sample could not be mounted directly on a tilt-stage, and so the tilting of the sample was restricted, and thus alignment of the plane of the film precisely with the plane of rotation had some deviations of up to 5$\deg$. This caused the reflected beam to precess slightly as the sample rotated. However, this could easily be compensated by replacing the collection lens with two confocal lenses. This scheme allowed placing the first of these lenses sufficiently close to the sample, ensuring a high enough numerical aperture for capturing the precessing reflected beam from the sample. We further stress that in our setup, and hence for all the images presented in this paper, the imaging was done by a regular convex lens and not by a microscope objective. Based on single element lenses, with different focal lengths, and by adjusting the object and image distances, we could easily control the resolution and the field of view. For the case when diffraction-limited resolution is needed, which was not necessary in our studies, where 5 - 10 \um  \ resolution was sufficient, the lens can be replaced by a high-magnification objective. Such high numerical aperture objective usually contains multiple optical elements that will have to be very close to the sample and, thus to the electromagnet, thus possibly causing a Faraday polarization rotation in the glass of the objective. Such an effect will then mix with the polarization rotation of the longitudinal MOKE signal. This problem was also observed with the short focal-length lenses that we have used. Nonetheless, this Faraday rotation induced in the lenses is precisely linear with the applied magnetic field, and thus could be filtered out digitally by looking at the slope of the measured intensity vs. $ H $ in the hysteresis loop at saturation. The calibration of the digital filter needed to be repeated only once per setup configuration.

	\section{Measurements}
\begin{figure*}
			\centering
			\includegraphics[width=\textwidth]{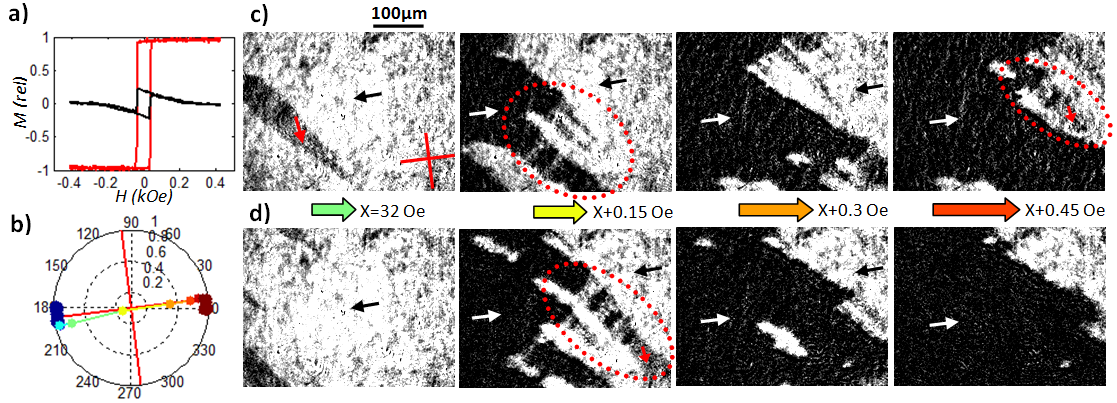}
			\caption{The reversal process for $ \<100\> $ Fe$ _{0.85}$Ga$ _{0.15} $ film on MgO substrate at $\alpha= 7\deg $ (angle between the easy axis and the applied field). (a) Hysteresis: longitudinal in red, transverse in black. (b) The vector magnetization computed from hystereses, with $ H $ given by the color-scale. $ \theta $ -- spin orientation, $ r $ -- spin magnitude ($ \ne 1 \Rightarrow$ multidomain state). (c, d) Each horizontal sequence shows one take in a real-time 30 fr/s movie, with the $ H $ field shown for each frame. The two sequences are taken under identical conditions. The red cross in the first frame indicates the orientations of the easy axes, and the black/red/white arrows -- local magnetizations. Contrast and size scales are the same for all images. We see that the double-step transition highlighted in imaging (c,d) is completely missed by the hysteresis (a,b).}
			\label{fig:100rev}
		\end{figure*}

	This setup allows for three different types of measurements. In the first configuration, the polarizer is set such that the beam arrives with s-polarization on the sample -- a geometry that can be used to measure the surface magnetization component parallel to the field based on the laser polarization rotation on reflection from the sample (longitudinal MOKE  \cite{domains_book}). To optimally detect this polarization rotation, we set the analyzer a few degrees from extinction, such that the derivative of the transmission curve (namely $ \sin^2(\theta) $ -- $ \theta $ being the analyzer angle) is significant, while the transmitted laser intensity is still low, giving the largest percent intensity variation. Further, to re-check the alignment of the polarizer and to minimize the residual effects of birefringence from the lenses used in the setup, we can set the analyzer exactly at extinction to verify that the resulting measured intensity remains independent of the magnetization. Because the slope of the transmission curve is zero at extinction, any observed dependence will indicate effects other than longitudinal MOKE such as transverse and higher-order MOKE signals, which can mostly be suppressed by adjusting the polarizer and analyzer alignment as well as their relative polarization axis orientation. Further, for the case of samples used in the present study, the resulting MOKE signal is strong enough to be clearly seen on the CCD in real-time during magnetic field reversal without any additional averaging or filters, other than background subtraction (see fig. \ref{fig:100rev}). We note here that the spatial and temporal resolutions in magnetic domain imaging are diffraction and frame recording rate limited, respectively. Note also that the time-resolution requirements can be somewhat relaxed by decreasing the field sweep rate, but not completely removed, as the observed reversal processes are dynamic and non-equilibrium. Finally, in this configuration, the photodiode simultaneously provides longitudinal MOKE hysteresis measurements with high temporal resolution of around 50 000 points/second.
	
	In the second setup configuration, we choose a p-polarized incident beam while removing the analyzer altogether, in order to measure the transverse MOKE magnetization component (orthogonal to the magnetic field), based on the reflectivity variations (transverse MOKE \cite{domains_book}). However, in such measurements the magneto-optical signal is much weaker, and imaging on the CCD requires additional data processing including temporal averaging, which precluded the use of real-time imaging. On the other hand, the photodiode detection is more sensitive and permits averaging, thus readily providing the spatially integrated transverse MOKE signal in real time. Nonetheless, if transverse MOKE imaging is needed, it can be achieved by redirecting the beam such that the plane of incidence is orthogonal to the horizontal scattering plane in fig. \ref{fig:setup} and to the sample. This geometry can then take advantage of the stronger signal of longitudinal MOKE for detecting the magnetization component perpendicular to the magnetic field direction and parallel to the sample plane. Finally, for the case of the samples studied here, the dominant contribution of the dipolar field confined the magnetization along the thin film plane and thus no significant contributions from the polar MOKE were detected.
	
	The possibility of switching between three different measurement configurations with our setup allows performing complementary measurements that can be used to simultaneously characterize the magnetic reversal in a sample. These setup configurations include the measurement of hysteresis loops of both components of the magnetization, as well as time-resolved longitudinal MOKE imaging of magnetic domains during the magnetization reversal. Such imaging is particularly interesting for studies of epitaxial ferromagnetic films that exhibit anisotropic magnetic hysteresis loops.  Moreover, simultaneous probing of the component-resolved hysteresis and magnetic domain structure during the reversal have allowed for unique identification of the magnetization orientation across the imaged region, ensuring complete visual information about the anisotropy of domain dynamics. 
	
	\subsection*{Example: two-step reversal from 4-fold anisotropy}
	In figure \ref{fig:100rev} we show the combined measurement capability of our setup, where $\alpha = 7\deg $ for a $ \<100\> $ Fe$ _{0.85} $Ga$ _{0.15} $ film -- the red and black loops illustrate respectively the longitudinal and transverse MOKE hysteresis, while the two horizontal image strips show two separate runs of the real-time domain dynamics, taken under identical conditions.\footnote{ We note that all images presented here are taken with a single exposure of the CCD (no averaging) and do not use any specialized filters besides the background subtraction.} The polar plot then shows the reversal of the vector magnetization, with the colormap indicating the external field $ H $, computed from the hysteresis. The points whose magnitude is less than 1 indicate multi-domain configurations corresponding to the shown images. 
	
	FeGa films have recently become particularly important due to their enhanced magnetostrictive properties \cite{FeGa_magnetostriction}, which can allow sensitive control of the film's magnetization using, e.g., thermal or electrical influences. We know from previous work in hysteresis\cite{dbl_step_hyst} and imaging \cite{MOIF2,MOKE90deg&spike}, that because the $ \<100\> $ films have two easy axes, the reversal usually proceeds via two $ 90\deg $ domain-mediated jumps. Our measurements have confirmed this behavior for this particular film for $ \alpha > 7\deg $. However, for $ \alpha<5\deg $, both hysteresis and imaging have shown a single $ 180\deg $-step reversal between initial and final states. Then, within the narrow transition region from one mode to the other, we have observed a curious mixture of the two pathways, which has not been reported previously, and is presented in fig. \ref{fig:100rev}. 
	
	In these images, we clearly see the presence of grey 90$ \deg $ domains, where the magnetization orientation is as indicated in the frames. These domains form either as the thin stripes emanating from the main black 180$ \deg $ domain, or as the characteristic grey and black bands over the triangular features emphasized in the images (fig. c,d). This behavior is, however, completely missed by the hysteresis measurements, which still indicate only a single $ 180\deg $-step jump, as seen from the straight, uninterrupted magnetization jumps measured (fig. a,b). This illustrates the necessity of imaging in the analysis of reversal pathways. Additionally, we stress that the observation of this mixture was only possible with the high contrast, spatio-temporal and angular resolutions provided by our MOKE imaging setup.

	\section{Procedures} \label{sec: meas}
	
	Next, we move on to elaborate on the measurement procedure. As illustrated by the two movie-strips in fig. \ref{fig:100rev}, the precise reversal dynamics and domain wall motion are different between the cycles of the external field, although after comparing the images vertically, we see that the overall structure and features of the reversal remain (this is also shown in \cite{real-time}). Thus, in order to sharply resolve the individual domain wall dynamics, real-time single-exposure imaging is necessary. Averaging over images obtained at the same applied field value in multiple field cycles (by triggering the CCD relative to the driving field) can still be useful to identify repeated reversal patterns, such as trap sites, but will smear out the important details of the wall motion seen in the figures here. Due to the practical restriction on the frame rate of the CCD (to 30 fr/s in our case), the real-time imaging requires a slowly varying external field near the coercivity value so as to decrease the domain wall speed and allow several frames to be recorded during the remagnetization process. Depending on the particular reversal pathway, the field sweep rates that we found practical were around 2 to 20 Oe/s. Although this can be achieved by simply decreasing the frequency of the sinusoidal seed from the signal generator, the resulting period can be up to one minute, with only about 10 frames containing non-trivial contrast information, resulting in extensive recording deadtime. This can be vastly improved (down to an acquisition period of around 5 s) by generating the seed on a computer such that the sweep rate is low only around the coercivity value, where it is needed. 
	
	The CCD is also connected to a PC, allowing to perform the background subtraction, while recording the resulting images in real-time, exactly as they are seen in the figures in this paper. The background image is chosen at the single-domain state at saturation, once at each new orientation of the magnetic field, and then subtracted from every measured frame in real time. We have seen that the system has enough stability that the background measurement need not be updated other than after system readjustments, while this method provides for easier interpretation of images than the common differential imaging technique (subtraction of subsequent frames).
	
	On the other hand, for hysteresis studies, particularly in the transverse configuration, averaging (over 5 to 30 loops) and large number of data points (such as anisotropy measurements of $  360\deg $ at $ 0.1\deg $ resolution) are often required. Since the temporal resolution of the photodiode is much higher than that of the CCD, it is convenient to increase the frequency of the driving field (to about 10 to 100 Hz), and hence the field sweep rate (to around 20 kOe/s). Although a correlation between the sweep rate and the reversal process has been reported in \cite{sweep_rate}, it is only manifested at frequencies above 100 Hz, and we have checked that in our system there were no significant differences between the hysteresis loops obtained at the two sweep rates. The signal from the photodiode is then fed into a computer, where and the data processing is performed, while the result is displayed live by the LabView software routine. Furthermore, our setup could then be automated to allow for fast and high angular resolution anisotropy measurements, thus allowing the detection of singular anisotropy features (e.g. spikes -- see below) that might otherwise be missed. To realize this at any given sample angle $ \alpha $, the field was cycled a pre-set number of times and the hysteresis averaged to obtain a clean measurement. The parameters of interest (such as coercivity and remanence) are then computed, and the rotation stage is subsequently rotated to the next specified angle, where the process is repeated. Automating this process typically allows completing an entire $ 360\deg $ anisotropy measurement at $ 0.1\deg $ resolution in 3 to 15 minutes depending on the precision requirements.\footnote{The LabView automation routine with a user guide is available from the authors upon request.} 
	
		\begin{figure*}
			\centering
			\includegraphics[width=\textwidth]{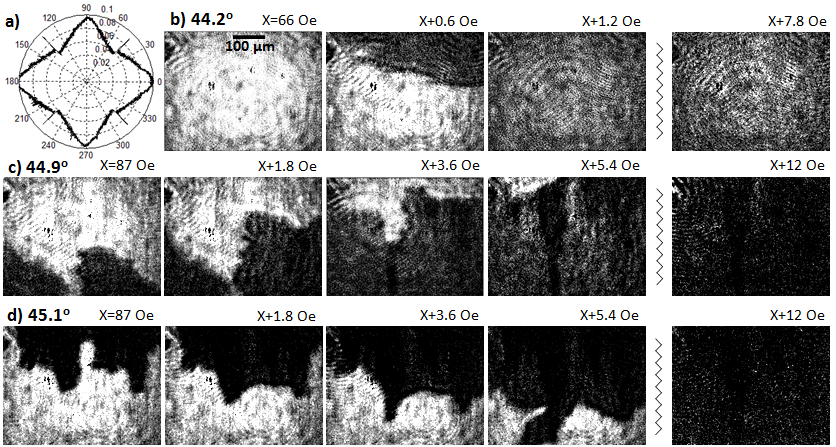}
			\caption{Same film as for fig. \ref{fig:100rev}. Observe the hard-axis spikes ($ <1\deg $ wide) in the coercivity anisotropy plot (a) (given in kOe). The three film-strips show the sharp distinction between the domain dynamics on (c,d) and off (b) the spike. The applied field is indicated above each frame, showing that the reversal proceeds much slower on the spike. The contrast is the same for all images, showing that only the first stage of the double-step transition is captured off the spike.}
			\label{fig:spike}
		\end{figure*}
	
	\subsection*{Example: hard axis coercivity spike from four-fold anisotropy }
	Figure \ref{fig:spike} shows one interesting scenario where high angular resolution is necessary to detect a distinct feature of the anisotropy -- the hard axis coercivity spike, which, despite notable interest over more than a decade (for example, see the hysteresis studies \cite{spike2003,spike2011}), is still not well understood. To study this, we have recorded the magnetic domain dynamics along the hard axis and slightly away from this direction in order to look at the mechanism behind the jump in the coercivity characteristic for the hard axis spike. In contrast to a prior report on the static magnetic domain imaging of domain structures near the hard axis direction \cite{MOKE90deg&spike}, our measurements could probe distinct differences in the magnetic domain dynamics, which could not be achieved previously. 
	
	Fig. \ref{fig:spike}a shows the coercivity anisotropy plot measured with the automated system as described above , where the coercivity of the hysteresis obtained at each angle is plotted. The aforementioned spikes, measuring $ <1\deg $ wide, are clearly observed at the hard-axes. The three film-strips then show how the domain behaviour changes going from off-spike (b) to on-spike on either side of the peak (c,d). The most dramatic change is in the domain wall speed -- as seen from the field step-size between the frames, the wall sweeps through the field of view much faster slightly away from the hard-axis spike (in fact, there, it is difficult to capture more than one frame in the multi-domain configuration).The average wall speed under otherwise identical conditions is about 15 times slower on the spike -- an observation which can only be detected with real-time imaging, again emphasizing its importance. Additionally, while away from the hard-axis spike the reversal proceeds through the usual double-step transition (with only the first step shown, as the coercivity of the second step diverges here), which is characterized by straight or zigzag domain walls (as in fig. \ref{fig:100rev}), along the hard-axis where the spike is observed, the reversal process is entirely different, resulting in amorphous walls whose shape is mostly governed by trapping behavior. Also note that the reversal process along the spike direction has a second stage, as seen by the remaining vertical triangular domains after the wall sweeps through. These domains then uniformly fade out, rather than moving through, thus marking another distinction from the usual double-step transition. Finally, we can also see that along the spike direction, the reversal dynamics are especially sensitive to the sample angle -- a rotation of only $ 0.2\deg $ results in the flip of the direction of domain wall propagation.

	\section{Conclusion}
	In conclusion, we have presented a cost-effective and easy to implement MOKE setup that is capable of directly imaging the time-resolved domain dynamics during the magnetization reversal in thin films. This was tested in particular for \FeGa films, and can easily be applied to other epitaxial ferromagnetic films. Our setup simultaneously allows for fast high-resolution anisotropy measurements based on the longitudinal and transverse MOKE hysteresis. Furthermore, we have illustrated the necessity of imaging, as well as the need for fine-tuning the anisotropy angle, in studies of the magnetic properties of epitaxial thin films. This was emphasized on several specific examples, where hysteresis measurements alone can miss crucial aspects of the reversal behavior, or even give misleading interpretations. We were also able to carry out real-time reversal imaging on the hard-axis coercivity spike, which should improve the understanding of such behavior in further studies.  
	
Our real-time MOKE microscopy approach allows to easily conduct anisotropy studies of surface domain dynamics, little of which has thus far been explored. As we have seen in a preliminary survey (which includes the above images), such studies can often lead to unexpected and interesting observations that can help deepen our understanding of magnetic domain dynamics in thin films.

	\bibliographystyle{unsrt}
	\bibliography{instr_ref}
	
\end{document}